\documentclass[12pt]{article}
\begin{document}
\title{John Bell Across Space and Time}
\author{Nino Zangh\`\i\ and Roderich Tumulka}
\date{July 7, 2003}
\maketitle

\begin{abstract}
  This is a review of the book \textbf{Quantum [Un]speakables: From
  Bell to Quantum Information}.  Reinhold A. Bertlmann and Anton
  Zeilinger (editors). xxii + 483 pp.  Springer-Verlag, 2002. \$89.95.
\end{abstract}

\textit{Ten years after his death, one of the sharpest minds in quantum
  physics was celebrated in a memorial conference.}

\bigskip

John Stewart Bell (1928-1990) was one of the leading physicists of the 20th
century, a deep and serious thinker. He worked at CERN in Geneva on the
physics of particle accelerators, made a number of impressive contributions to
quantum field theory, and became famous for the discovery of a phenomenon he
called nonlocality. However, the most remarkable thing about him was perhaps
that he was a realist.

Realism is the philosophical view that the world out there actually exists, as
opposed to the view that it is a mere hallucination. We are all born realists,
but some of us change our minds as adults. Now it may seem to you that for
physics to make any sense, a physicist would have to be, or at least pretend
to be, a realist; after all, it would seem that physics is about finding out
how the world out there works.

But, as a matter of fact, in the 1920s Niels Bohr, the leading quantum
physicist of his time, began to advocate the idea that realism is childish and
unscientific; he proposed instead what is now called the ``Copenhagen
interpretation'' of quantum physics, a rather incoherent philosophical
doctrine, which (according to Richard Feynman) ``nobody really understands.''
Part of this doctrine is the view that macroscopic objects, such as chairs and
planets, do exist out there, but electrons and the other microscopic particles
do not.  Correspondingly, Copenhagen quantum theory refuses to provide any
consistent story about what happens to microscopic objects, and instead
prefers to make contradictory statements about them. According to the
Copenhagen view, the world is divided into two realms, macro and micro,
``classical'' and ``quantum,'' logical and contradictory---or, as Bell put it
in one of his essays, into ``speakable'' and ``unspeakable.''

Although it is not clear where the border between the two realms should be,
and how this duality could possibly be compatible with the fact that chairs
consist of electrons and other particles, Bohr's view became the orthodoxy.
That is, it became not merely the majority view among physicists, but rather
the dogma. Ever since, being a realist has been rather dangerous for a quantum
physicist, because it has been widely regarded as a sign of being too stupid to
understand orthodox quantum theory---which, as we've mentioned, nobody really
understands.

Along with Albert Einstein, Erwin Schr\"odinger, Louis de Broglie and David
Bohm, Bell was one of the few people who felt compelled by his conscience to
reject Bohr's philosophy. Bell emphasized that the empirical facts of quantum
physics do not at all force us to renounce realism: There is a realist theory
that accounts for all of these facts in a most elegant way---Bohmian mechanics
(also known as de~Broglie--Bohm theory). It describes a world in which
electrons, quarks and the like are point particles that move in a manner
dictated by the wave function. It should be taught to students, Bell insisted,
as a legitimate alternative to the orthodoxy.  And in 1986, GianCarlo
Ghirardi, Alberto Rimini, and Tullio Weber succeeded in developing a second
kind of realist theory, encouraged by Bell and known as \emph{spontaneous
  localization}. But overcoming prejudice and changing convictions takes more
than one generation.

\textit{Quantum [Un]speakables} is the proceedings volume of a conference held
at the University of Vienna in November 2000 to commemorate the 10th
anniversary of Bell's death. The 30 articles written for this volume by 35
authors deal foremost with nonlocality and, of course, the meaning of quantum
theory. The contributions focus very much on personal recollections and mostly
presuppose that the reader is familiar with the relevant physics and
mathematics. The recollections make this book a valuable source both on John
Bell the man and on the history of quantum physics between 1950 and 1990.
Among other things, several authors complain about the dogmatic aversion among
physicists in the 1960s to even take note of Bell's nonlocality theorem.

\textit{Quantum [Un]speakables} also reflects the prevailing situation in the
year 2000 in that it collects personal, diverging views about the meaning of
quantum physics from a cross-section of physicist. The cross-section is
biased, though, because researchers working on Bohmian mechanics, of which
Bell was the leading proponent during the decades before his death, were
simply not invited to the conference, and the realists are in the minority
among the authors. Thus we recommend that readers be very cautious in regard
to the conclusions drawn in this book about the foundations of quantum
physics.

This warning concerns in particular the conclusions drawn from Bell's
nonlocality theorem. Let us tell the story briefly here. Bohmian mechanics
involves superluminal action-at-a-distance and thus violates the ``locality
principle'' of relativity theory. This was considered, by the Copenhagen camp,
an indication that Bohmian mechanics was on the wrong track. In 1964, Bell
proved that any serious version of quantum theory (regardless of whether or
not it is based on microscopic realism) must violate locality. This means that
if nature is governed by the predictions of quantum theory, the ``locality
principle'' is simply wrong, and our world is nonlocal. It also means that the
nonlocality of Bohmian mechanics is not a sign of its being on the wrong
track, but quite the contrary.

The Copenhagen view, in comparison, is indeed less local: It is nonlocal in
cases that Bohmian mechanics can explain in a purely local way. (For example,
for a particle in a quantum state that is a superposition of being in London
and being in Tokyo, according to Copenhagenism there is no matter of fact
about whether the particle actually is in London or in Tokyo prior to the
first attempt at detection---which presupposes a temporal ordering.)  But it
is also contradictory, vague and confusing enough for its adherents to claim
it is completely local, and thus that nonlocality is a consequence of an
attachment to \textit{realism}.  Therefore, so the argument goes, it was Bell
who finally proved realism wrong!  Bell, of course, emphatically rejected this
incorrect interpretation of his nonlocality theorem.

The crucial experiments violating Bell's inequality and thus, according to
Bell's theoretical analysis, demonstrating nonlocality have been performed
many times since 1980, and have also lead to significant improvements in
experimental techniques. Some of these techniques have now become valuable
for quantum cryptography and the first steps towards the construction of a
quantum computer. These two fields are usually summarized under the key
word "quantum information," and great hopes are expressed, also in
\textit{Quantum [Un]speakables}, that quantum information will provide new
insights into the nature of the quantum world. 

But we see no reason for such hopes. Quantum information theory is a
straightforward application of the rules laid down in, for example, John von
Neumann's classic 1932 book on the mathematical foundations of quantum
mechanics. Any interpretation of quantum mechanics, to the extent that it
succeeds in explaining these rules, also explains quantum computers and the
like.  And to the idea that quantum theory may after all be merely about
information and nothing else, Bell responded with a crucial question:
``Information? Whose information?  Information about what?''

\bigskip

\noindent \textbf{Reviewer information.} \textit{Nino Zangh\`\i\ is professor
  of theoretical physics at the Universit\`a degli Studi di Genova, Italy.
  Roderich Tumulka is a post-doctoral research fellow at the physics
  department of the Universit\`a degli Studi di Genova. The authors wish to
  thank Sheldon Goldstein for his critical reading of a draft for this
  article.}

\end{document}